\documentclass{appolb}
\usepackage{epsfig}
\usepackage{amsmath,amssymb}

\newcommand{\beqn}{\begin{equation}}
\newcommand{\eeqn}{\end{equation}}
\newcommand{\req}[1]{Eq.\,(\ref{#1})}

\begin{document}
\title{Strong Field Physics:  Probing Critical Acceleration \\
and Inertia with Laser Pulses and Quark-Gluon Plasma%
\thanks{Joint\,\,report\,\,of\,\,individual\,\,presentations\,\,made\,\,by\,\,both\,\,authors\,\,at\,\,the\,\,50\,\,Cracov\\School\,\,of\,\,Theoretical\,\,Physics\,\,held\,at\,Zakopane,\,\,Poland\,\,June\,\,9-19,\,\,2010}
}
\author{Lance Labun and Jan Rafelski
\address{Department of Physics, The University of Arizona, Tucson, 85721 USA}
}
\maketitle
\begin{abstract}
Understanding physics in domains of critical (quantum unstable) fields requires investigating the classical and quantum particle dynamics at the critical acceleration, $\dot u \to 1$ [natural units].  This regime of physics remains today experimentally practically untested.  Particle and light collision experiments reaching critical acceleration are becoming feasible, in particular applying available high intensity laser technology.  Ultra-relativistic heavy ion collisions breach the critical domain but are   complicated by the presence of much other physics.  The infamous problem of radiation reaction and the challenging environment of quantum vacuum instability arising in the high field domain signal the need for a thorough redress of the present theoretical framework. 
\end{abstract}

\section{Introduction}
Strong applied `external' fields have long been understood to probe physical properties of the quantum vacuum state.  Considerable effort was committed to exploration of vacuum decay by positron production in scattering of large $Z$ nuclei~\cite{Muller:1972zz} at sub-atomic scale. Though motivated by a different set of questions, the study of ultra-relativistic heavy nuclei collisions and the associated multi-particle production explores the physics of strong fields in quantum chromodynamics~\cite{Casher:1978wy}.  Slowly-varying $(\lambda \gg m^{-1})$ external fields of arbitrary strength are employed as a formal tool in theoretical investigations to gain insight into non-perturbative structures of quantum theories of matter interactions~\cite{Dunne:2004nc}.  The questions posed when these approaches combine are: what happens when a  particle   collides with a ultra strong, slowly-varying field~\cite{Fedotov:2010ja}\,?  Is our theoretical framework able to cope with this domain of physical parameters~\cite{Rafelski:2009fi}\,?

The question is more subtle than it may at first appear, because it probes the foundations of the physical concepts of inertia and acceleration.  Ernst Mach noted the intricate connection between inertia and acceleration and that the understanding of acceleration requires introduction  of an additional inertial reference frame~\cite{MachReviewR,MachReviewB}.  This widely accepted insight is independent of Mach's proposal to relate inertia to the matter content of the Universe, which is in disagreement with experiment~\cite{MachReviewR}.  With a few important exceptions (e.g.~\cite{Brans:1961sx}), the necessity of the additional reference frame in the framework for accelerated motion has received little attention in past 100 years of modern physics, perhaps since the magnitude of acceleration  we have considered has been `small.'  We will show that experiments probing `strong' acceleration are feasible today.

\begin{figure}
\centerline{
\includegraphics[width=0.7\textwidth]{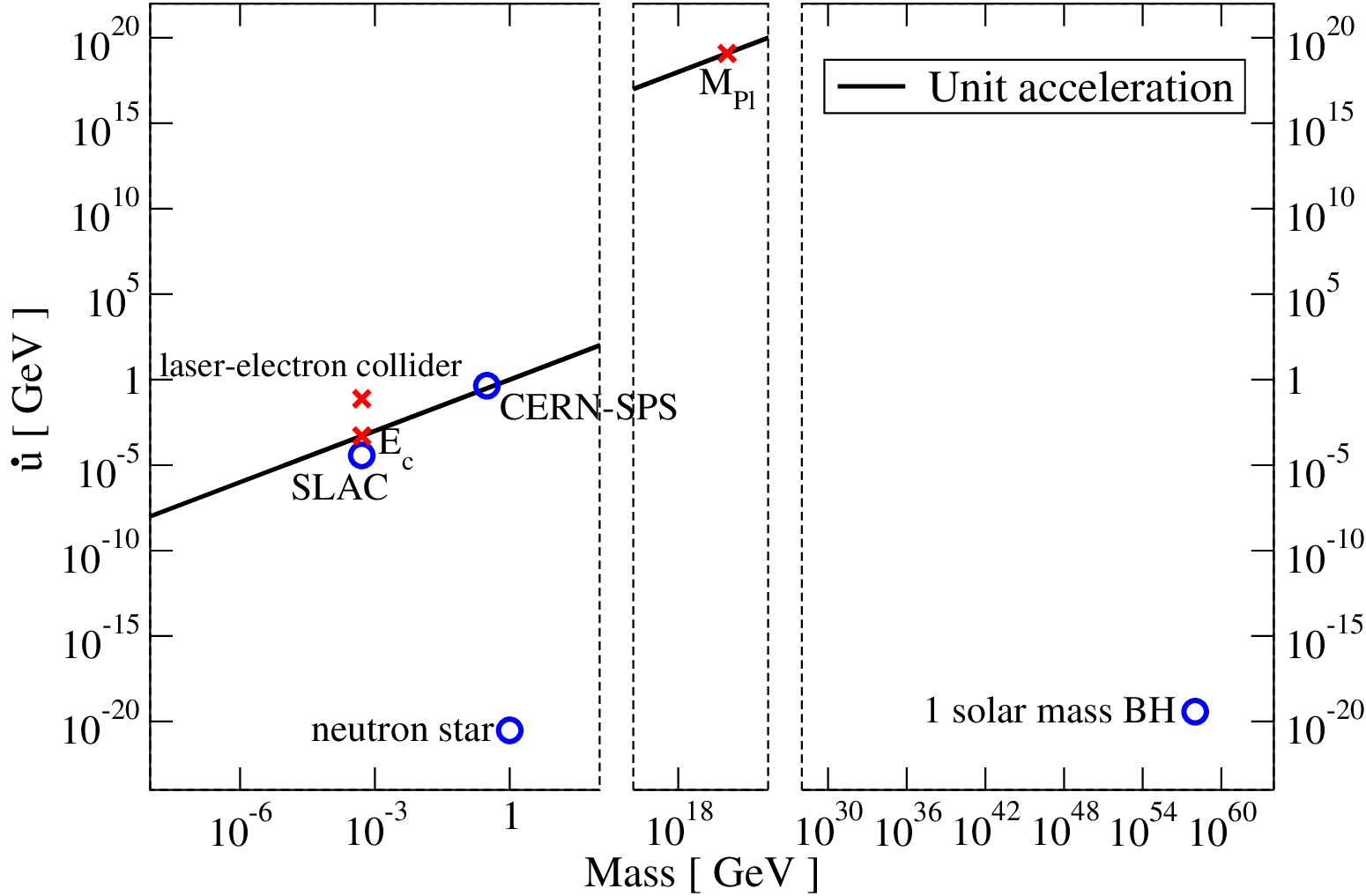}
}
\caption{ Circles mark accelerations presently available, either by laboratory experiment or in the universe.  The $X$s highlight unattained accelerations.  The surface gravity of the Earth is too below the scale of the chart to include.
\label{ma-scatter}}
\end{figure}

Consider the least massive particle readily available to experiment, an electron, entering an extended space-time domain where the electromagnetic field nears the `critical' strength
\beqn\label{Ec-defn}
E_c = \frac{m_e^2 c^3}{e\hbar} = 1.32\times 10^{18}~{\rm V/m}=4.41\times 10^9c~{\rm T},
\eeqn
with $m_e$ the electron mass. According to the Lorentz force equation, 
\beqn\label{Lorentz}
\frac{du^{\alpha}}{d\tau}=-\frac{e}{m}F^{\alpha\beta}u_{\beta},
\eeqn
this is equivalent to the condition 
\beqn\label{acrit}
\left|\frac{du^{\alpha}}{d\tau}\right| \to \frac{m_ec^3}{\hbar}\ \equiv 1\,[m_e] =2.33\times 10^{29} \:\frac{\rm m}{\rm s^2}.
\eeqn
For comparison, the acceleration of a proton at the surface of a $1.5M_{\odot}$, 12 km radius neutron star is `only' $g_{\rm surf} \simeq 1.39 \times 10^{11}\:{\rm m/s}^2$.  Figure~\ref{ma-scatter} is a scatter plot showing different physical domains and offers a feeling for the relative scales.  

Neither classical, nor the related quantum electrodynamic (QED) description can be believed to be complete as $E$ approaches $E_c$---the quantum theory because the electromagnetic field becomes unstable to spontaneous pair creation~\cite{Sauter:1931zz,Schwinger:1951nm}, and the classical theory because  radiation reaction becomes non-negligible~\cite{Dirac:1938nz}. The looming possibility of performing experiments involving critical fields $E/E_c \to 1$~\cite{Mourou:2006zz} thus requires a fresh look at charged particle classical and quantum dynamics for the case of an applied critical field.  

We survey here the physics associated with the unit-acceleration regime. We will show how the challenge of describing particle dynamics at this scale will lead to deeper understanding of the building blocks of quantum and classical theories, due particularly to the need to render their descriptions consistent and complete.  In the wider view encompassing the present theory of acceleration, namely general relativity, the creation and experimentation with large-scale strong fields $E \sim E_c$ will have ramifications for any unified description of particle interactions.

We first recall in section~\ref{Laser}  the  experimental advances of laser pulse technology. In subsection~\ref{Lcollider} we discuss a light pulse-electron collider as makes an experiment with macroscopic strong fields possible in the foreseeable future.   We then recall in section~\ref{aa} the conceptual context of (critical) acceleration  and present a thorough discussion of radiation reaction and its connection to the understanding of inertia. We connect in section~\ref{RHIC} to the acceleration reached in high energy elementary and heavy ion collisions. In section~\ref{quantum} we address  the appearance of critical acceleration in quantum field theories and the related vacuum instability.  We briefly compare the physics of accelerated observers and accelerated vacuum as it presents itself at this time in section~\ref{aether}.

\section{High-Intensity Lasers}\label{Laser}
\subsection{Technological limits}
Renewed motivation for addressing   strong field physics comes from advances in laser technology.  The technique of chirped pulse amplification~\cite{Mourou:2006zz} has made possible laser pulses of higher contrast, shorter length, and relativistic intensities where the field amplitude $e|A|$ is many times the electron rest mass.  To surpass the exawatt $10^{22}$ W/cm$^2$ record intensities already achieved~\cite{Yanovsky:2008}, concerted efforts, such as the Extreme Light Infrastructure with intended sustained laser pulse intensity as high as $10^{24}$ W/cm$^2$, are in final planning stages~\cite{ELI}.  

In field strengths, these intensities respectively correspond to
\beqn
|E|= 2.7\times 10^{15}-10^{16}~ {\rm V/m,} 
\eeqn
or in terms of the dimensionless, normalized amplitude commonly used 
\beqn
a_0 \equiv \frac{e|A|}{m_ec}= \frac{e|E|}{m_e\omega c} = 51-510
\eeqn
at optical frequencies $\omega \sim 1$\:eV/$\hbar$.  As the technical challenge grows with the intensity, proportional to $a_0^2$, laser technology faces still some effort to bridge the remaining few orders of magnitude to $a_0\simeq mc^2/\hbar \omega\to 5\times 10^5$ corresponding to the critical field $E_c$.

Along side increasing average intensity, fine-tuning the control over the shape of laser pulses proceeds apace.  Pulses consisting of only 10 waves are possible today, which translates into a duration
\beqn
10\lambda = 10\frac{hc}{\rm 1\:eV} = 12.5 \:\mu{\rm m} = 40 c\:{\rm fs}.
\eeqn
Contrast at the front of the laser pulse controls much of the pulse-particle interaction since for these extraordinarily high intensities the prepulse can already alter in a dramatic way the state with which the main laser pulse interacts.  Developments such as relativistic plasma mirrors~\cite{Wu:2010ul} have produced dramatic improvements in contrast, approaching the physical limit on the rise of the potential, $\lambda/4$, imposed by the wavelength of light used.  Since force is proportional to the gradient of the potential, this limit implies that for an electron to achieve unit acceleration in a laser field, the wavelength---as observed by electron---must be decreased.

\subsection{Laser pulse-electron collider}\label{Lcollider}
To confront a particle with unit force, we therefore propose to take advantage of the Lorentz transformation of the electromagnetic field, in a manner analogous to the particle colliders such as LHC and RHIC.  An electron with momentum $\vec p = \gamma m\vec v$ incident on a laser pulse with the momentum (wave-vector) $\vec k$ will see the field Doppler-shifted in its rest frame
\beqn\label{Doppler}
\omega \to \omega'=\gamma(\omega+\vec v \cdot \vec k).
\eeqn
The force and acceleration achieved in the collision is thus multiplied by $\sim 2\gamma$ for nearly opposite momenta.  The difficulty in arranging the collision also increases with $\gamma$, because the pulse appears contracted
\beqn
l_{\Vert}'=l_{\Vert}(\gamma \cos \theta)^{-1}, \quad l_{\perp}'=l_{\perp}(\gamma\sin\theta)^{-1}
\eeqn
in the longitudinal and transverse directions, respectively.  

Some of the technical challenges have been discussed in the literature and a laser pulse-electron storage ring, analogous to those in hadron-hadron or electron-electron machines, has been proposed~\cite{Huang:1997nq}.   The recent advances in laser technology have reinvigorated investigation of how to observe the QED nonlinearities arising as $E\to E_c$~\cite{Marklund:2008gj}.  Yet even before laser intensities climbed rapidly in the past decade, the capability to perform experiments of this sort was demonstrated by \cite{Burke:1997ew,Bamber:1999zt}, and thus we know that it is possible to arrange a head-on collision of a relativistic 47\,GeV electron beam with a high intensity focused light pulse.  The peak normalized laser intensity attained 15 years ago was 
\beqn
a_0 = 0.4
\eeqn
and with $46.6$ GeV incident electrons consequently saw an acceleration
\beqn
\dot u  \le 0.073\,[m_e].
\eeqn
While this magnitude falls short of critical $\dot u \to 1$, the experiment successfully observed  multi-photon phenomena predicted by strong field QED, including evidence for nonlinear Compton scattering and electron-positron pairs produced via a Breit-Wheeler process.  Repeating the SLAC experiment with 15 years' improved off-the-shelf laser technology would be a relatively easy step
and would lead to study of physics arising at critical acceleration.

The center-of-mass energy available in a laser pulse-electron collision surpasses that commonly available in other major probes of strong field physics, including hadron-hadron collisions studied at RHIC (see section \ref{RHIC}).  A laser pulse consisting of 
\beqn
n\omega \geq 10^{22}\times 1\:{\rm eV~~photons} = 1.6\:{\rm kJ}
\eeqn
is a coherent object of mass-energy $10^{10}$ times larger than the TeV protons available at the LHC.
Today it is possible to imagine a coherently-synchronized 1000-beam system (NIF has 192) with individual beam intensities 10 times higher. Such a futuristic hyper-laser would still comprise only a percent-fraction of the Planck mass
\beqn\label{Mpl}
M_{\rm Pl} = 1.22 \times 10^{28}\:{\rm eV}~ 
 \approxeq 10^6 \times (1.6\:{\rm kJ~laser~pulse}).
\eeqn
Although achieving Planck energy remains far away, coherent light pulses appear much closer to this goal than any other known physical system and certainly offer much better hope than hadron-hadron colliders~\cite{Casher:1997rr}.

\section{Acceleration and Radiation Reaction}\label{aa}
\subsection{Planck scale acceleration}
We have set the stage to argue that experiments can be performed today at unit acceleration, which is a natural scale to expect new physics.  Recall that `unit value' quantities defined by setting $G=\hbar=c=1$ are Planck's units that arise from the closed, natural dimensioned system created by the introduction of $\hbar$ (see the last page in his work~\cite{Planck}).  Unit acceleration in natural units $\dot u \to 1[m]$, thus presents the Planck scale of the `theory of acceleration.'  In order to avoid fruitless discussions of the matter, we do not call critical acceleration `Planck acceleration' since this clearly was not on the list he presented.

Acceleration, viewed as a change in 4-velocity in a given field, is different for different mass particles. The condition \req{acrit} is made universal normalizing by the mass, introducing the critical specific acceleration $\aleph$
\beqn \label{unitsa}
\aleph\equiv \frac{\dot u}{m} \to \frac{c^3}{\hbar} = 1.37\times 10^{23}\frac{\rm m}{\rm s^{2}\,kg}.
\eeqn
Newtonian gravity near a body of Planck mass $M=M_{\rm Pl}\equiv\sqrt{\hbar c/G}$ exhibits unit acceleration at Planck distance $L_{\rm Pl}=\hbar/(M_{\rm Pl}\,c)$,
\beqn
\aleph_G\equiv \frac{\dot v}{M_{\rm Pl}} = \frac{G}{r^2} 
\to \frac{c^3}{ \hbar} ~~{\rm at}~~r=L_{\rm Pl}.
\eeqn
Creating elementary, or coherent, Planck mass (energy) scale objects and probing the gravity-related Planck scale directly will remain a challenge for some time to come.  However, the Equivalence Principle assures that we access the same physics when realizing critical acceleration \req{unitsa} in the context of other (electromagnetic and/or strong) interactions.
 
\subsection{Is  electromagnetism a consistent theory?}
Consider a localized electron incident on a large $n$ laser field.  The classical character of the initial state leads one to believe that classical dynamics as encoded by the Maxwell equations and the Lorentz force~\req{Lorentz} should apply.  However, it has long been recognized that the system of classical equations of motion is incomplete~\cite{Dirac:1938nz}.  Specifically, the inhomogeneous Maxwell equation
\beqn\label{Max-inhomo}
\partial_{\beta}F^{\beta\alpha} = j^{\alpha}
\eeqn
leads to radiation by an accelerated charge at the rate
\beqn\label{rad-rate}
\mathcal{R}= \frac{dE}{dt_{\rm lab}}
   =-\frac{2}{3}e^2\left(\!\frac{du^{\alpha}}{d\tau}\!\right)^{\!2}
\eeqn
(see for example~\cite{Jackson}).  The inertial reaction of the particle to this energy-momentum loss is not included in the Lorentz force~\req{Lorentz}.  This observation leads to the iterative procedure:
\begin{enumerate}
\item Solve Lorentz force for spacetime path of particle in prescribed external field.
\item Compute emanating radiation fields according to Maxwell equations.
\item Correct field configuration and return to step 1.
\end{enumerate}

Efforts to close the iteration loop into a single dynamical radiation-reaction incorporating equation have occupied authors from Lorentz through the present.  The earliest effort, the Lorentz-Abraham-Dirac equation~\cite{Abraham:1902},
\beqn \label{LADeqn}
m\dot u^{\alpha}=-eF^{\alpha\beta}u_{\beta}+m\tau_0\left(\ddot u^{\alpha}-u^{\beta}\ddot u_{\beta}u^{\alpha}\right),
\eeqn
was shown by Dirac to follow from covariant energy-momentum conservation in the action \req{EMaction}~\cite{Dirac:1938nz}. However, \req{LADeqn} fails to be a predictive equation of motion, because it contains third derivatives of position $\ddot u$, requiring a third boundary condition to specify a solution.  While the extra condition is able to eliminate unstable `run-away' solutions of \req{LADeqn}, its utilization is generally regarded as an unphysical violation of causality.

It is of importance to realize that in principle the consistency of classical electromagnetism is not assured, as there are several rather independent dynamical components;  the total action   is the sum of electromagnetic, matter, and interaction components
\beqn
\label{EMaction}
\mathcal{I} 
=-\int d^4x \frac{1}{4}F^2 +
\frac{mc}{2}\int_{\rm path}\!d\tau \,(u^2-1) + q\int_{\rm path} \!dx^{\alpha}(\tau) A_{\alpha}.
\eeqn
The latter two integrations follow the spacetime path of the particle with charge $q$ ($=-e$ for the electron).  The equations of motion for the total system are usually obtained separately for the field and the particle: Maxwell's \req{Max-inhomo} by variation of the 1st and 3rd terms, and the Lorentz force \req{Lorentz} by variation of the 2nd and 3rd terms.  Combining the two into one dynamical equation \req{LADeqn} exhibits their inconsistency, since the derivation of \req{LADeqn} appears to be physically sound, yet does not produce a physical theory, except in the limit when the acceleration is small $\dot u \ll m$.  In this case, the radiation rate is also small.  In particular, since (using natural units)
\beqn
\dot u \sim mE/E_c \quad {\rm and} \quad
  \mathcal{R}\sim e^2(\dot u)^2=e^2\left(\frac{mE}{E_c}\right)^2,
\eeqn
the effect of radiation reaction can be treated perturbatively,  as long as $E/E_c < 1$.  A rigorous procedure for carrying this expansion to arbitrary order has been recently derived~\cite{Gralla:2009md}.  However, a deeper understanding of the physics is necessary to be able to make predictions effectively in the unit-acceleration regime.

\subsection{Efforts to improve the classical theory}
Many efforts in the intervening years have attempted other solutions (see~\cite{Rafelski:2009fi} for a list), most recently even including some quantum effects~\cite{Sokolov:2010jx}.  Of these, the equation set forth in~\cite{LandauLifshitz} has received the most attention, because it implements the iteration procedure described above into a perturbative expansion by replacing
\beqn
\ddot u^{\alpha}\to \frac{d}{d\tau}\left(-\frac{e}{m_e}F^{\alpha\beta}u_{\beta}\right).
\eeqn
The resulting nonlinear equation, confirmed by the expansion in~\cite{Gralla:2009md}, 
\beqn\label{LLEq}\begin{split}
m\dot{u}^{\alpha} &= -e F^{\alpha\beta}u_{\beta}-e\tau_{0}\Big\{F_{,\gamma}^{\alpha\beta}u_{\beta}u^{\gamma}
-\left.\frac{e}{m}\left[F^{\alpha\beta} F_{\beta\gamma} u^\gamma - F^{\beta\gamma} F_{\gamma\delta}u^\delta u_{\beta} u^\alpha\right]\right\},\\[0.2cm]
&\hspace*{3cm}\tau_0 = \frac{2}{3}\frac{e^2/4\pi\varepsilon_0}{mc^3}=6.26\times  10^{-24}\:{\rm s},
\end{split}\raisetag{0.8cm}\eeqn
can be solved analytically in several simple but useful prescribed fields, including notably a transverse `laser' electromagnetic wave~\cite{DiPiazza,Hadad:2010mt}. 

This  solution  of \req{LLEq} exhibits several features important for potential experiment. It displays damping of the electron motion, as would be expected from the dissipative nature of radiation, and consequently predicts less total integrated radiation emission than the uncorrected Lorentz force.  This prediction, among others, demonstrate that the effects of radiation reaction will be easily discernible in the radiation emission and trajectories of the electrons.  In the case of a head-on collision between a highly relativistic electron and a high intensity laser pulse, an electron Lorentz factor $\gamma=10^3$ and normalized laser intensity $a_0=100$ more than suffice to make radiation reaction effects visible~\cite{Hadad:2010mt}.  

An important point to recognize is that the trajectories predicted by the Eqs.\,\eqref{LADeqn} and \,\eqref{LLEq} are  different even when both are considered valid.  This distinction arises from the perturbative expansion producing the Landau-Lifshitz expression~\req{LLEq}, which converges more weakly as the radiation correction terms become comparable to the Lorentz force itself, and hence approximately as
\beqn\label{LLRRcond}
\omega a_0^{2}\gamma \tau_0 \sim 1
\eeqn 
where $\gamma$ is the Lorentz factor of the incident electron~\cite{LandauLifshitz,Hadad:2010mt,Koga2}.  This condition differs from the expected onset of the relevance of quantum effects
\beqn\label{Qonset}
\omega a_0\gamma \sim m_e
\eeqn
corresponding to critical acceleration and the critical field strength $E_c$, \req{Ec-defn}, observed in the electron's rest frame.

\subsection{Accessibility of radiation-reaction effects in classical domain}
\begin{figure}
\centerline{
\includegraphics[width=0.75\textwidth]{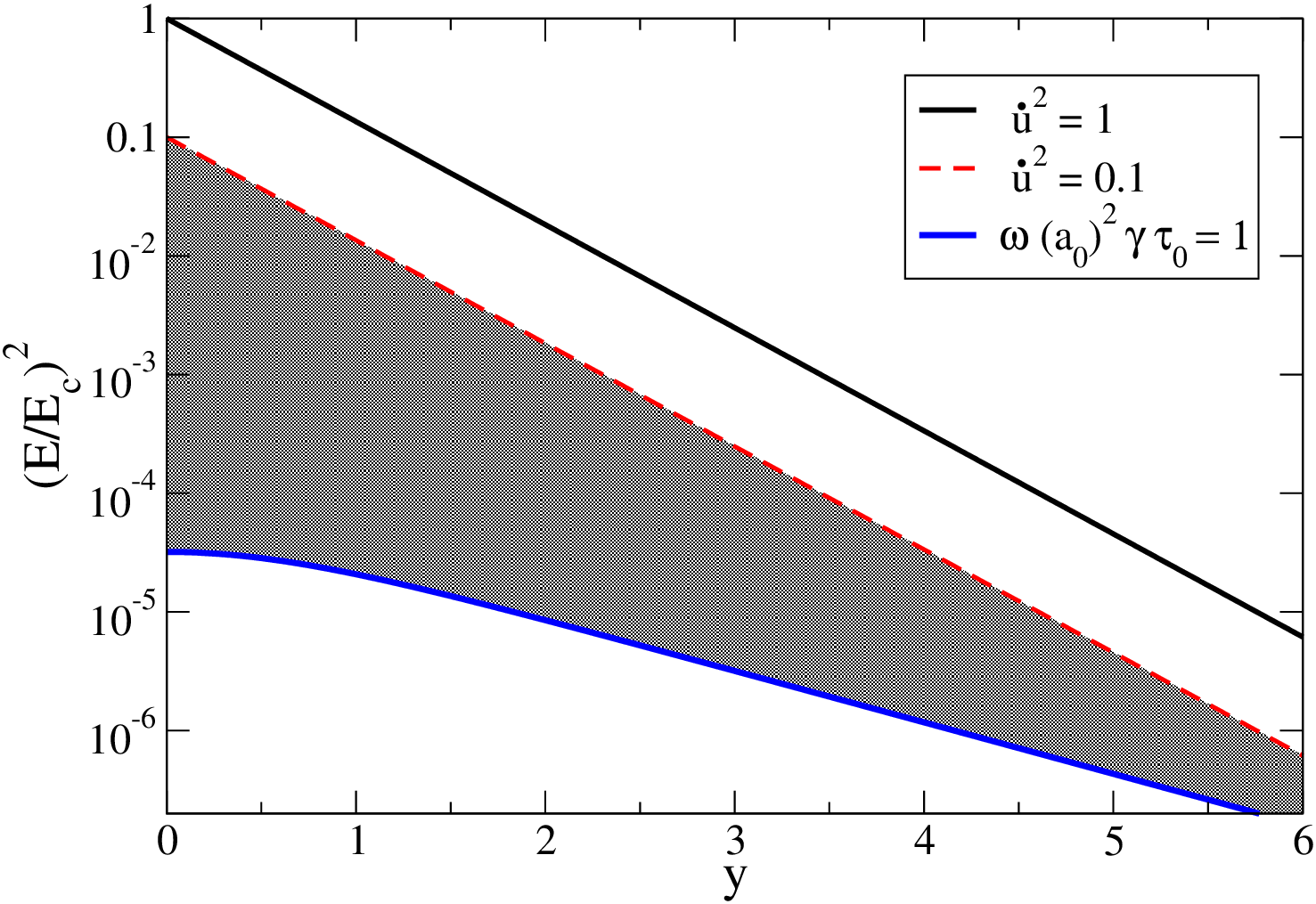}
}
\caption{Domain of interest for electron of rapidity $y$ moving oppositely to a laser field of normalized intensity $(E/E_c)^2$. The Lorentz-force critical (unit) acceleration is the upper solid line. The shaded region delimited by \req{Qonset} from below and $a <0.1$ from above highlights the parameter space where classical dynamics are expected to remain dominant and radiation reaction effects will be non-negligible.
\label{RRcontours}}
\end{figure}

The accessibility of radiation reaction effects within the domain of classical dynamics can be seen directly by considering the invariant spacelike acceleration
\beqn\label{AAA}
a^2=-\dot u^{\alpha}\dot u_{\alpha} 
\eeqn
for an electron colliding with a laser place wave according to the Lorentz force~\req{Lorentz}. We consider an electron of rapidity $y$, 4-velocity 
$$u^{\alpha}=(\cosh y,0,0,-\sinh y)$$, 
incident on an oppositely traveling   wave
\begin{align}
u^{\alpha}F_{\alpha\beta}&=(\cosh y,0,0,-\sinh y)\left(\begin{array}{c c c c} 0 & E_1 & 0 & 0 \\ -E_1 & 0 & 0 & B_2 \\ 0 & 0 & 0 & 0 \\ 0 & -B_2 & 0 & 0 \end{array}\right)  \nonumber\\[0.4cm]
 &= (0,E_1\cosh y+B_2\sinh y,0,0) = (F_{\beta\alpha}u^{\alpha})^{\rm T} \nonumber
\end{align}
and thus
\begin{align}
a_L^2 &= -\frac{e^2}{m^2}(u^{\alpha}F_{\alpha\beta}F^{\beta\gamma}u_{\gamma})
 = \frac{e^2}{m^2}(E_1\cosh y + B_2\sinh y)^2  \\[0.4cm]
&=\frac{e^2}{m^2}\left\{\left(\!\frac{E_1\!+\!B_2}{2}\!\right)^{\!2}e^{2y}+\left(\!\frac{E_1\!-\!B_2}{2}\!\right)^{\!2}e^{-2y}+\frac{E_1^2\!-\!B_2^2}{2}\right\}
 \to \left(\frac{eE}{m}\right)^{\!2}e^{2y} \notag
\end{align}
The contours in figure~\ref{RRcontours} display the levels at which radiation reaction effects are important according to \req{LLRRcond} and the onset of quantum effects at unit acceleration, \req{Qonset}.  We allow considerable margin for the quantum effects to become relevant due to the sensitivity of a real experiment to residual matter triggering cascades~\cite{Fedotov:2010ja} and frequency effects~\cite{Hebenstreit:2009km}. The shaded region highlights parameter space where classical dynamics are expected to remain dominant while radiation reaction effects are non-negligible, specifically requiring an improvement over \req{LLEq}, due to the poor convergence of the Landau-Lifshitz expansion. 

Radiation reaction must therefore be dealt with  also in a regime where classical dynamics are dominant.  Investigations such as~\cite{Hadad:2010mt,Koga2,DiPiazza2} begin to address the gap in understanding high acceleration dynamics within classical theory.  An open question is how much guidance quantum electrodynamics (QED) will offer in this endeavor. There is widespread belief that QED has cured defects of classical theory, yet there is no firm evidence that this is the case.  At this point it has been shown~\cite{Higuchi:2005an,Higuchi:2009ms} that QED can reproduce the classical equations of motion to second order in perturbation theory when unusual boundary conditions are imposed.  The issues preventing understanding of the relationship of time-reflection invariant QED and a causal classical theory have been recognized often in the past but not resolved, see comments made by Spohn in Section 1 of his recent monograph~\cite{Spohn:2004ik}.

\section{Elementary  and  Heavy Ion High Energy Collisions}\label{RHIC}
\subsection{Critical acceleration and quark-gluon plasma formation}
An interesting early precursor to the current discussion of high acceleration phenomena is the hypothesis by Barshay and Troost~\cite{Barshay:1977hc} that achieving high acceleration in elementary interactions could explain the thermal characteristics of the multiparticle production phenomena in terms of the Hawking-Unruh effect (see section~\ref{quantum}).  This idea has been further elaborated in recent years by Satz and collaborators~\cite{Satz:2008zza,Castorina:2007eb}.  We do not take the  perspective that all multiparticle production is due to strong acceleration effects: there are excellent reasons within the realm of strongly interacting matter to expect the emergence of a thermalized state in heavy ion collisions. However, in high energy particle collisions the strong acceleration phenomena could contribute and thus help establish thermal equilibrium of particle yields.

To illustrate this line of thought, we first show that some matter  participating in heavy ion reactions does achieve critical acceleration due to the strong stopping of quarks that has been reported.   While the larger fraction of each nucleus passes through without large loss of rapidity (energy), a significant fraction of valence quarks,  at the level of $\simeq 5\% $ of both projectile and target are stopped in the center of momentum frame~\cite{Stopping}. This means that the scaling domain, expected at ultra-high energy~\cite{Bjorken:1982qr}, has not been reached. For initial momenta of the components $M_i$ 
\beqn
p^{\mu}=\left( M_i\cosh y_p,0,0,\pm M_i\sinh y_p\right),
\eeqn
where $y_p$ the rapidity of the incident beam, the acceleration $a$, \req{AAA}, required to stop a parton within a proper time $\Delta \tau$ is
\beqn
a=\dot y\simeq\frac{y_p}{\Delta \tau}.  
\eeqn
$y_p=5.4$ at RHIC and $y_p=2.9$ at CERN-SPS, so for a constituent quark of mass $M_i\simeq M_N/3\simeq 310$~MeV to undergo critical acceleration, it must be stopped within $\Delta \tau< 3.4$~fm/c or $\Delta\tau<1.8$~fm/c, respectively at RHIC and SPS.   In fact, for the SPS experiments with a 30 GeV beam incident on a fixed target, $\Delta \tau < 1.3$ fm/c.  This time scale is comparable to the `natural' quark-gluon plasma formation time $\tau_0=0.5-1$~fm/c~\cite{Bjorken:1982qr}.

\subsection{Anomalous soft photon production}
In addition to approaching critical acceleration with respect to quantum chromodynamic forces,  high energy particle collisions reveal the possibility of incomplete understanding of radiation reaction in the high energy regime. Many of the colliding particles and in particular the strongly-interacting quarks carry electromagnetic charge.  Their high accelerations during the collision generate gluon and photons.  The effect of radiation reaction due to this radiation accentuates the stopping power and augments the gluon and photon radiation yield.  Such considerations could explain the puzzling excess of soft photon production above perturbative QED expectation in many experiments. 

Wong offers a comprehensive summary of the phenomena in need of an explanation~\cite{Wong:2010gf}.  Anomalous soft photon production in elementary high energy interactions is observed universally in conjunction with the production of hadrons, mostly mesons: in $K^+ p$ reactions~\cite{Chl84,Bot91}, in $\pi^+ p$ reactions~\cite{Bot91}, in $\pi^- p$~\cite{Ban93,Bel97,Bel02a}, in $pp$ collisions~\cite{Bel02}, in high-energy $e^+$-$e^-$ annihilations through $Z^0$ hadronic decay~\cite{DEL06,DEL08,DEL09,Per09}.

The main features of the anomalous soft photon production are summarized as follows:
\begin{enumerate}
\item
Anomalous soft photons are produced in association with hadron
production at high energies.  They are absent when there is no hadron
production \cite{DEL08}.
\item
The anomalous soft photon yield is proportional to the hadron yield.
\item
The anomalous soft photons carry significant transverse momenta, in the range
of many tens of MeV/c.
\item
The yield of anomalous soft photons increases approximately linearly with
the number of neutral or charged produced particles.
\end{enumerate} 
Especially the first and last feature suggest that photons and gluons are produced together in strong stopping of quarks, with gluons turning into neutral hadrons at hadronization.  This corroborates the possibility that the quark stopping is driven by effective radiation reaction forces, akin to those we discussed for the case of  the classical electromagnetism. To prove this conjecture will require a more complete study of the radiation reaction phenomena at the quark level.

\section{Quantum Vacuum and Acceleration}\label{quantum}
\subsection{Vacuum Instability} 
As noted by Sauter~\cite{Sauter:1931zz}, Euler and  Heisenberg~\cite{EKH}, and Schwinger~\cite{Schwinger:1951nm}, strong electric fields are susceptible to conversion into electron-positron pairs. The field strength $E_c$, \req{Ec-defn} is the non-perturbative  scale of the barrier to vacuum decay.  The materialization is global and very rapid when the field achieves critical strength---i.e. when electrons and positrons experience above-critical acceleration.  

A practical description of the decay is as a semi-classical tunneling process, which becomes non-negligible as the potential becomes strong enough to accelerate an electron across the gap in the Fermi spectrum,
\beqn
\frac{eE}{m/\hbar c}=\frac{\rm Slope~of~potential}{\rm Scale~of~wavefunction} \sim {\rm Gap~width}=2mc^2 
\quad {\rm or}\quad 
E \sim \frac{2m^2c^3}{e\hbar}. \notag
\eeqn
Up to a factor 2, this is just the condition for critical field strength~\req{Ec-defn} and unit acceleration~\req{acrit}.

\begin{figure}
\centerline{
\includegraphics[width=0.75\textwidth]{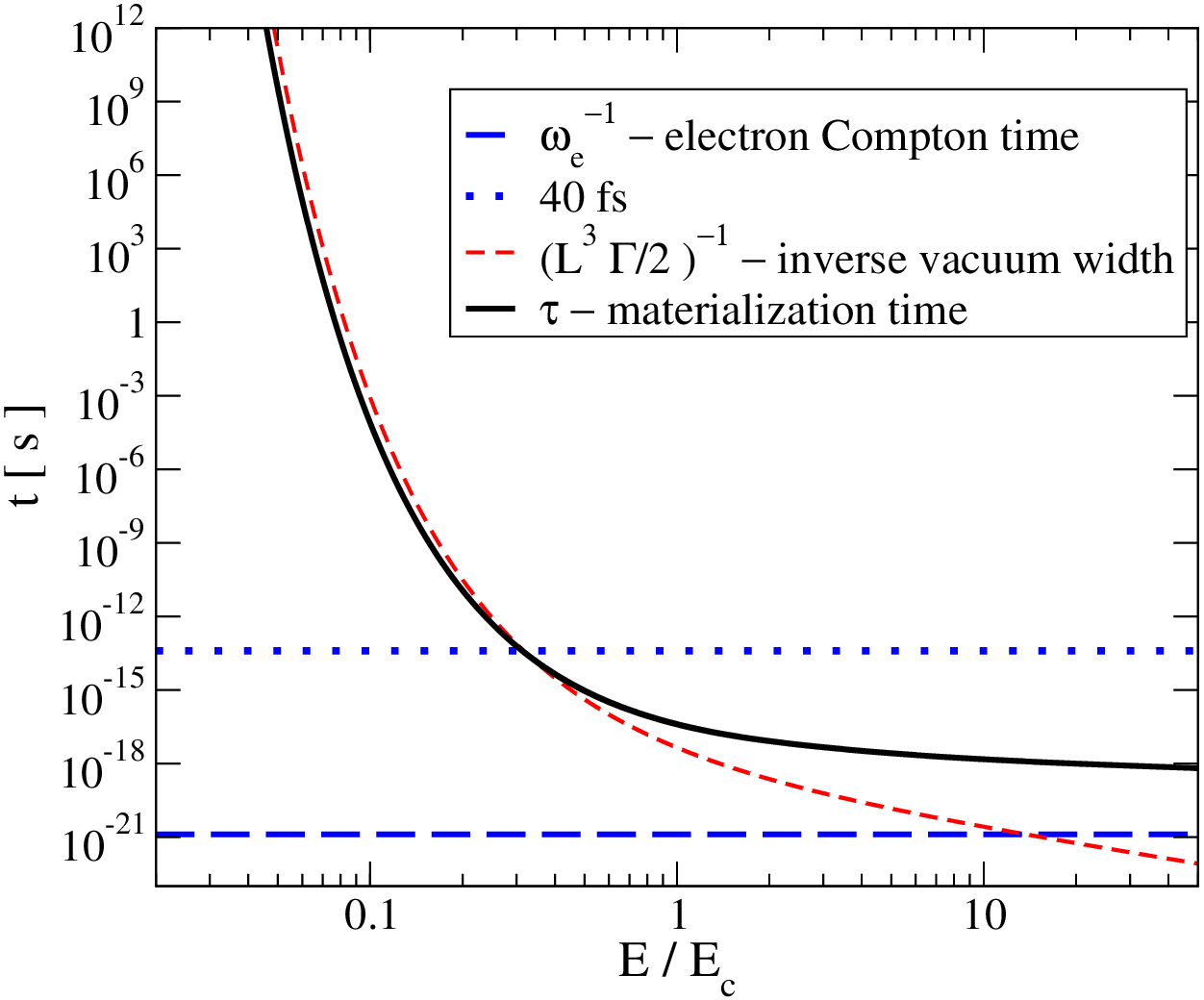}
}
\caption{The lifespan (inverse decay half-width) of the with-field quantum vacuum state (dashed line) and the energy content of the field (solid line) as function of the externally applied field strength $E/E_c$.  The time length of presently available laser pulses (40 fs, short-dashed horizontal line) and the inverse Compton frequency of the electron (long-dashed horizontal line) provide reference for the time scale.
\label{fig:stability}}
\end{figure}

Two typical timescales are associated with the lifespan of the quantum vacuum state with applied field. 
\begin{enumerate}
\item  The total probability of decay via pair-creation of the zero-particle state in the presence of a given field strength,
\beqn\label{ImagV}
\Gamma = \frac{(eE)^2}{4\pi^3}\sum_{n=1}^{\infty}\frac{1}{n^2}\exp\left(\!-n\frac{ \pi E_c}{E}\!\right),
\eeqn 
is the `width' of the with-field state and hence the imaginary part of an effective potential. This rate does not illuminate what happens to the state at strong field but simply conveys the message how quickly the instability takes hold. This is the dashed (red) line  in figure~\ref{fig:stability}.
\item
A physical definition of the persistence of the field is obtained by studying the conversion of field energy into particle pairs~\cite{Labun:2008re}. As the critical field is approached, the vacuum materializes at a rate 
\beqn\label{tauvac}
\tau^{-1} = \frac{1}{u_f}\frac{d\langle u_m\rangle}{dt},
\eeqn
where $u_f$ is the energy density available in the electromagnetic field and 
\beqn
\frac{d\langle u_m\rangle}{dt} = \frac{2_s eE}{4\pi^2}\int_{m_e}^{\infty}\!dM_{\perp}\:2M_{\perp}^2 e^{-\pi M_{\perp}^2/E}
\eeqn
is the expected rate at which energy is converted into electron-positron pairs. The result is shown as the solid line in figure~\ref{fig:stability}. 
\end{enumerate}

For comparison the inverse electron Compton frequency (long-dashed horizontal line) and the typical laser pulse time, 40 fs (short-dashed horizontal line) are also shown in figure~\ref{fig:stability}. Interestingly, we note that the laser pulse lifespan is of the same magnitude as the pulse length already at $E=0.3E_c$.  On the other hand the materialization of the field energy is not as fast as the Compton frequency.  This result is due to a factor $\alpha/\pi$ difference between $\omega_e^{-1}$ and the dimensionful coefficient in $\tau$ and can be interpreted to mean that the weakness of the QED coupling implies that it is not necessary to implement back-reaction of produced pairs on the applied field, though such an approach is certainly required for a fully consistent description~\cite{Kluger:1992gb}.

\subsection{Subcritical pair production}
The analytic completion of the imaginary part \req{ImagV} is the effective potential derived by Euler, Heisenberg and Schwinger,
\beqn\label{Veff}
V_{\rm eff} = i\Tr\ln (G^{-1}[F_{\rm ext}]) = 
 -\frac{2_s}{16\pi^2}\int_0^{\infty}\!\frac{dt}{t^{3+\epsilon}}\left(\frac{eEt}{\tan eEt}-1\right)e^{-m^2t}.
\eeqn 
The Green's function in the external field $G[F_{\rm ext}]$ appearing in \req{Veff} can be decomposed as a sum over eigenstates in the external field
\beqn\label{Greensfn}
G(x,x') = 
\sum_{\lambda=\{p^{\mu},\sigma\}}c_{\lambda}\,\hat T \psi_{\lambda}(x)\bar\psi_{\lambda}(x'),
\eeqn 
where $\psi_{\lambda}$ is an eigenfunction of the with-field Dirac equation, $\hat T$ the time-ordering operator, and $c_{\lambda}$ a normalization constant.  This propagator has been explicitly computed \cite{Nikishov:1964zza,Eberly:1969} and was used for the baseline predictions for the electron-laser pulse collisions at SLAC~\cite{Burke:1997ew,Bamber:1999zt}.  As noted the experiment remained below the level at which quantum effects are important and critical acceleration is attained, and relatively good agreement seen between the semi-classical evaluation of $G(x,x')$ utilized and the predominantly classical conditions in the setup.

\subsection{Accelerated vacuum and Hawking-Unruh radiation} 
The decomposition of the Green's function \req{Greensfn} shows how $V_{\rm eff}$ integrates the effect of the external field on the single particle states and thus represents an `accelerated' vacuum state.  
The natural next question is whether this description of the vacuum accelerated by the external field is consistent with the description of the quantum vacuum according an accelerated observer.
 
Unruh showed~\cite{Unruh:1976db} that a detector (a scalar particle confined to a box) undergoing constant acceleration $a$ displays a Planckian excitation spectrum with Bose statistics and temperature
\beqn\label{Thu}
T_{\rm HU} = \frac{a}{2\pi}.
\eeqn
Subsequent work has re-derived this result from many different approaches, and in every case, the temperature agrees with the Hawking-Unruh temperature and the statistics match those of the un-accelerated quantum field~\cite{Crispino:2007eb}.  

In contrast, the effective action \req{Veff} has a quasi-Planckian representation
\beqn\begin{split}
V_{\rm eff} =& \frac{m^4}{(4\pi)^2} \frac{\pi T_{\rm EH}}{m}\!\int_0^{\infty}\! 2_s\ln(\omega^2/m^2-1+i\epsilon)\ln(1-e^{-\omega/T_{\rm EH}})d(\omega/m),\\
&\hspace*{3cm} T_{\rm EH} = \frac{eE}{\pi m}=\frac{a}{\pi},
\end{split}\eeqn
in which the temperature is twice the Hawking-Unruh temperature, $T_{\rm EH}=2T_{\rm HU}$, and the statistics of the distribution are inverted, displaying a Bose-like ($-$)~\cite{Muller:1977mm}.  Computing the effective potential $V_{\rm eff}$ for a spin-less ``electron'' results in the same temperature $T_{\rm EH}$ and a `wrong' Fermi-like sign in the distribution~\cite{PauchyHwang:2009rz}.  This situation is summarized in Table~\ref{ARtable} and reveals puzzles that remain in the non-perturbative predictions of the quantum theory at the critical scale $E\to E_c$.

\begin{table}
\caption{The thermal characteristics of acceleration radiation contrasted with those found for a constant electric field.\label{ARtable}}
\begin{center}
\begin{tabular}{c|c}
Accelerated Observer & Accelerated Vacuum \\
\hline\hline
detector accelerated against & electron Fermi sea states 
\\ flat-space vacuum &  at constant acceleration $a=eE/m$ 
\\ \hline
detector response function & sum negative energy states \\
$\rightarrow$ thermal excitation spectrum & $\rightarrow$ effective potential\\
$ T_{\rm HU} = a/2\pi$
& $ T_{\rm EH} = a/\pi$
\\ \hline
statistics match & statistics inversion \\
({boson} $\mapsto$ {boson})  & ({boson} $\mapsto$ {fermion}) \\
({fermion} $\mapsto$ {fermion})  & ({fermion} $\mapsto$ {boson})\\ 
\hline\hline
\end{tabular}
\end{center}
\end{table}

\subsection{The Quantum Vacuum Frame and the \AE ther}\label{aether}
One only appreciates the challenge of bringing together the discussion of acceleration and the quantum vacuum recollecting the conflicting descriptions of accelerations provided by general relativity and quantum theories.  First, one should bear in mind that in the absence of quantum theory, extended matter objects  are hard to imagine. Without a finite size, point-like particles fall freely in gravitational fields, and there is no acceleration to be discussed.  Resistance to gravitational free fall is in essence only possible since quantum atoms have finite size and many atoms come together to form macroscopic objects that resist the pull of moderate gravitational forces.  In turn, the ability to construct devices producing critical acceleration originates in the quantum nature of matter and radiation.  We may therefore expect that the study of physical phenomena at unit acceleration is likely to advance our understanding of the difficulties in uniting gravitational and quantum theories, since their inconsistency is accentuated and the equivalence principle challenged. 
 
Further, quantum theory contains an universal inertial state, which is the global, lowest-energy (i.e. quantum vacuum) state. Textbook treatments of the classical limit of quantum theory do not involve this reference frame. However, in presence of strong acceleration a more refined classical limit could be required which generates a modified Lorentz equation, wherein acceleration refers to the vacuum state as the universal inertial frame.  A promising method to derive a classical limit that includes vacuum dynamics as well as back-reaction has been developed within the relativistic Wigner function  formulation~\cite{BialynickiBirula:1991tx}.

Within this framework of phase space functions there seems to be a natural opportunity to derive particle dynamics with the vacuum state present as a reference frame identifying classical acceleration. The classical dynamical equations in phase space arise naturally~\cite{Rafelski:1993uh}; the classical limit requires coarse-graining and so far back-reaction effects have not been considered. The effective forces obtained were at this level of discussion identical to the known classical Lorentz and Bargmann-Michel-Telegdi equations. 

Noteworthy in the above discussion is that the quantum vacuum assumes the role of the relativistically invariant \ae ther, the intangible carrier of physical law that Einstein proposed around 1920~\cite{Einstein}, reexamining his earlier criticism  of the \ae ther hypothesis.  This modern \ae ther (i.e. quantum vacuum) provides a natural preferred inertial frame perhaps capable of resolving the debate inspired by Mach and Einstein over how to define inertia.  The challenge of understanding inertia may thus require the inclusion quantum vacuum structure.

Remarkably, the current paradigms of quantum field theory invoke quantum vacuum structure as the preeminent source of the definition of the inertial mass of all particles, from electroweak symmetry breaking and minimal Higgs coupling to color confinement.  On the other hand, reconnecting the presence of the quantum vacuum to the classical realm remains difficult with the classical limit of quantum theory eluding the full understanding, and in particular not referring acceleration to the presence of the vacuum state.

\section{Conclusion}\label{conclude}

Taken in isolation, many of the physics topics discussed here are well-known, if not in every case perfectly understood.  Our purpose has been to unite apparently disparate phenomena and in-principle considerations under the common theme of high-acceleration physics and point to a few new insights, some at present still hypothetical, but all accessible to in depth study via experimental technologies either immediately available or presently in development.

Laser technology nears the capability to perform finely-controlled experiments in which charged particles attain and considerably exceed the critical acceleration $\dot u \to 1\,[m]$.  Since the quantum vacuum structure of electrodynamics is simpler than that of chromodynamics, we expect that this context will provide cleaner experimental and theoretical access to the physics of radiation reaction and dynamics in the unit-acceleration regime.  On the other hand, the lesson of heavy ion collisions and multi particle production may be that the phenomena are fundamentally linked by critical acceleration.

To pose, much less answer the question of defining acceleration, the  dynamical theory of matter and radiation must at least incorporate the inertial response of the charge to its own radiation, a self-consistency not yet obviously in hand for classical or quantum electrodynamics.  Thus, while investigation of particle production and quantum vacuum structure may offer guidance in the quest to understand matter and inertia, the associated challenges in the classical domain, and in particular radiation reaction, must be independently addressed.

The ongoing and forthcoming experimental efforts will without doubt lead to  a renaissance in strong field physics.  The array of topics covered here and the connection through the Equivalence Principle to the gravitational theory and Planck scale highlights both difficulties and opportunities provided by strong fields and critical acceleration to understand the present theoretical framework encompassing quantum and gravitational theories.\\[0.4cm]

\noindent This work was 
supported by  the grant from
the U.S. Department of Energy, DE-FG02-04ER41318. 


\end{document}